\documentstyle[twoside,fleqn,jhuwkshp,psfig,epsfig,color]{article}

\newcommand{\xmm}{{\it XMM-Newton} }
\newcommand{\chandra}{{\it Chandra} }

\newcommand{\rxte}{{\it RXTE} }

\def\lapp{\ifmmode\stackrel{<}{_{\sim}}\else$\stackrel{<}{_{\sim}}$\fi}

\def\gapp{\ifmmode\stackrel{>}{_{\sim}}\else$\stackrel{>}{_{\sim}}$\fi}
\def\spose#1{\hbox to 0pt{#1\hss}}

\def\approxlt{\mathrel{\spose{\lower 3pt\hbox{$\sim$}}
        \raise 2.0pt\hbox{$<$}}}
\def\approxgt{\mathrel{\spose{\lower 3pt\hbox{$\sim$}}
        \raise 2.0pt\hbox{$>$}}}

\begin{document}

\title{The Chandra High Energy Transmission Grating Spectrometer  probes the {\sc dusty warm absorber} in the Seyfert 1 galaxy MCG--6-30-15 }
 
\author{J.C. Lee,  
\address{MIT Center for Space Research, 77 Massachusetts Ave., Cambridge, MA. 02139 }
C.R. Canizares,$\rm ^a$
H.L. Marshall,$\rm ^a$ A.C. Fabian,$\rm ^b$
R. Morales,$\rm ^b$ 
N.S. Schulz,$\rm ^a$
K. Iwasawa \address{Institute of Astronomy, Madingley Road, Cambridge, CB3 OHA, U.K.}}

\begin{abstract}
The \chandra HETGS spectra of the Seyfert~1
galaxy MCG--6-30-15  show numerous narrow, unresolved 
(FWHM $\approxlt$~200 $\rm km \,s^{-1}$) absorption lines  from a
wide range of ionization states of N, O, Mg, Ne, Si, S, Ar, and Fe.
The initial analysis of these data, presented in Lee et al. (2001), 
shows that a dusty warm absorber 
model adequately
explains the spectral features $\approxgt$ 0.48~keV ($\approxlt$ 26 \AA ).
We attribute previous reports of an apparently  highly redshifted O~{\sc vii }
edge  to the neutral Fe~L   absorption complex and 
the O~{\sc vii } resonance series (by transitions higher than He~$\gamma$; 
He~$\alpha,\beta,\gamma$ are also seen at lower energies). 
The implied dust
column density needed to explain the Fe~{\sc i}~L edge feature 
agrees with that obtained from earlier
reddening studies, which had already concluded
that the dust should be associated with the ionized absorber (given the
relatively lower observed X-ray absorption by cold gas). Our findings 
contradict the interpretation of
Branduardi-Raymont et al. (2001), based on \xmm RGS spectra,
that this spectral region is dominated by highly relativistic soft X-ray
 line emission originating near the central black hole.  Here we review these
issues pertaining to the soft X-ray spectral features as addressed by Lee et al., (2001).
{\tt Details found in Lee et al., 2001, ApJ., 554, L13  \cite{lee01}} 
\end{abstract}

\maketitle

\section{Introduction}
Recently, workers analyzing the \xmm RGS data of the luminous ($\rm L_X
\sim 10^{43} \rm erg \, s^{-1}$) nearby ($z$=0.0078) Seyfert 1 galaxy 
MCG--6-30-15 and the similar object Mrk~766 have proposed a radical alternative to the
warm absorber model as the origin of the spectral features in the soft
($\approxlt$ 2~keV) band (Branduardi-Raymont et al. 2001 \cite{br2001}, hereafter BR2001).
It had been generally accepted that the features were imposed by
partially ionized absorbing material at $\approxgt$ parsec distances
from the black hole. The main signatures  are strong  O~{\sc vii} and 
O~{\sc viii} absorption edges, which are nearly
ubiquitous in Seyfert~1 galaxies.
However,  BR2001 proposed that in MCG--6-30-15 and Mrk~766 the observed
spectral features are  soft X-ray emission lines from close to the
black hole  highly broadened by relativistic effects on top of a 
very flat continuum.  This scenario
was invoked to explain the apparent 16,000~$\rm km \, s^{-1}$ redshifted O~{\sc vii}
edge without  associated resonance lines.

Our \chandra High Energy Transmission Grating Spectrometer (HETGS) observation does
not support this interpretation -- it shows that a warm absorber
model, with the addition of a contribution from dust, can describe the
spectrum of MCG--6-30-15.  The main points of the
arguments presented in Lee et al., (2001) \cite{lee01} are reiterated here, and
the reader is referred to that paper for details.   The MCG--6-30-15
luminosity, assuming $\rm H_0 = 50 \,\rm km\,s^{-1}\,Mpc^{-1}$ is
$L_{\rm x} \sim 1.3 \times 10^{43}$~erg/s for the flux state reported
in our paper.  In contrast, the flux of MCG--6-30-15
during the epoch of the \xmm look presented by BR2001 is
comparable to the `deep minimum' studied by Iwasawa et al., (1996) \cite{i96}.
The best fit Galactic ($N_H \sim 4 \times 10^{20} \, \rm cm^{-2}$)
absorbed power-law to the \chandra data is $\Gamma \sim 1.9$ consistent 
with simultaneous \rxte observations (e.g. Fig.~1). 

\begin{figure}[t] % fig.1
%\vspace{10pt}
\centerline{\psfig{file=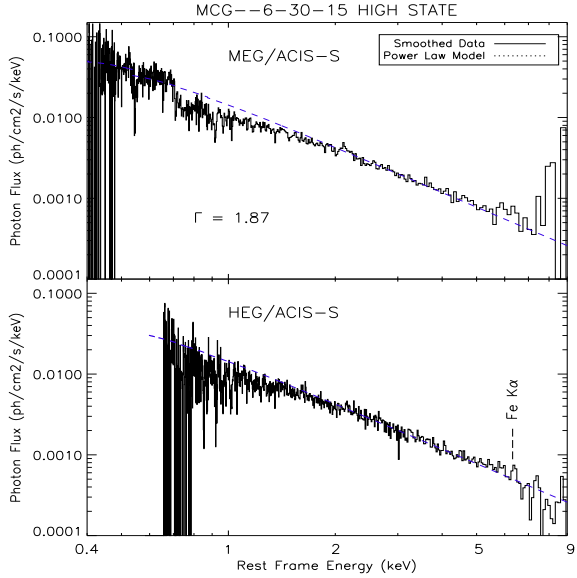,width=3.0in,height=3.0in}}
\vspace{-0.3in}
\caption{A simple power-law of $\Gamma \approx 1.9$ modified by 
Galactic ($N_{\rm H} = 4.06 \times 10^{22} \, \rm cm^{-2}$) absorption fits the
MEG and HEG data well.  The 0.7-2.5~keV region are excluded from our 
fits in order to mitigate the effects of excess absorption in that 
energy range.  The 6.5$-$8~keV iron region is also excluded. }\label{fig-Gpl}

\end{figure}                                                                             

\begin{table}[t]   
\begin{center}
\begin{tabular}{lccc} 
\multicolumn{4}{c}{\small \sc O~{\sc vii} and O~{\sc viii} resonance lines in MCG--6-30-15}\\ \\
\hline
\hline
\hline \\
\multicolumn{4}{c}{\small \color[rgb]{1,0,0} O~{\sc vii : }  \sc $\tau > 0.2$ or $N_{O \sc {vii}} > 7 \times 10^{17} \rm cm^{-2}$, and $b \sim 100 \rm \,km\, s^{-1}$} \\ 
\\
\multicolumn{4}{c}{\small \color[rgb]{0,0,1} $\tau_{O\sc vii : } \sim 0.6-0.8$ (at 0.74~keV edge) ==$>$  $N_{O \sc {vii}} > 2.5 \times 10^{18} \rm cm^{-2}$}
\vspace{0.2in} \\
\hline
\hline
\\
{\small \em \rm Species} & {\small \em \rm $\lambda$ }  & {\small $^\dagger f_{ij}$ } & {\small \rm EW } \\
\small O~{\sc vii} $ 1s^2 -$ & \small   {\AA} ({\rm eV}) &  & \small m\AA (eV)\\
\\
\hline
\hline \\
\small $ 1s2p\;$ (He~$\alpha$)& \small 21.6019 (574)  &\small  $6.96 \times 10^{-1}$ & $17 \pm 7$ (0.45) \\ \\
\small $ 1s3p\;$ (He~$\beta$) & \small 18.6288 (666)  &\small  $1.46 \times 10^{-1}$ & $^a$ see below\\ \\
\small $ 1s4p\;$ (He~$\gamma$) & \small 17.7680 (698)  &\small  $5.52 \times 10^{-2}$ & $13 \pm 2$ (0.51)\\ \\
\small $ 1s5p\;$ (He~$\delta$) & \small 17.3960 (713)  &\small  $2.68 \times 10^{-2}$ & $^b$ see below\\ \\
\small $ 1s6p\;$ & \small 17.2000 (721)  &\small  $1.51 \times 10^{-2}$ & $13 \pm 2$ (0.54)\\ \\
\small $ 1s7p\;$ & \small 17.0860 (726)  &\small  $9.37 \times 10^{-3}$ & $11 \pm 2$ (0.47)\\ \\
\small $ 1s8p\;$ & \small 17.0092 (729)  &\small  $6.22 \times 10^{-3}$ & $12 \pm 2$ (0.51)\\ \\
\small $ 1s9p\;$ & \small 16.9584 (731)  &\small  $4.34 \times 10^{-3}$ & $13 \pm 2$ (0.56)\\ \\

\hline 
\hline 
\hline \\
\multicolumn{4}{c}{\small Table 1: $\dagger$ oscillator strength, $^a$ confused -- near N~{\sc vii}  edge, }\\
\multicolumn{4}{c}{\small $^b$ confused -- near  Fe~L~{\sc iii}, L~{\sc ii} edge}
\end{tabular}
\end{center}
\label{tab-oxyres}
\end{table}
\vspace{-0.2in}

\section{Evidence for Warm Absorption :  Dust vs. GR}

\begin{figure}[h] % fig.2
%\vspace{10pt}
%\centerline{\psfig{file=lee_jc_fig2.eps,angle=0,width=3.5in,height=3.0in}}
\epsfig{file=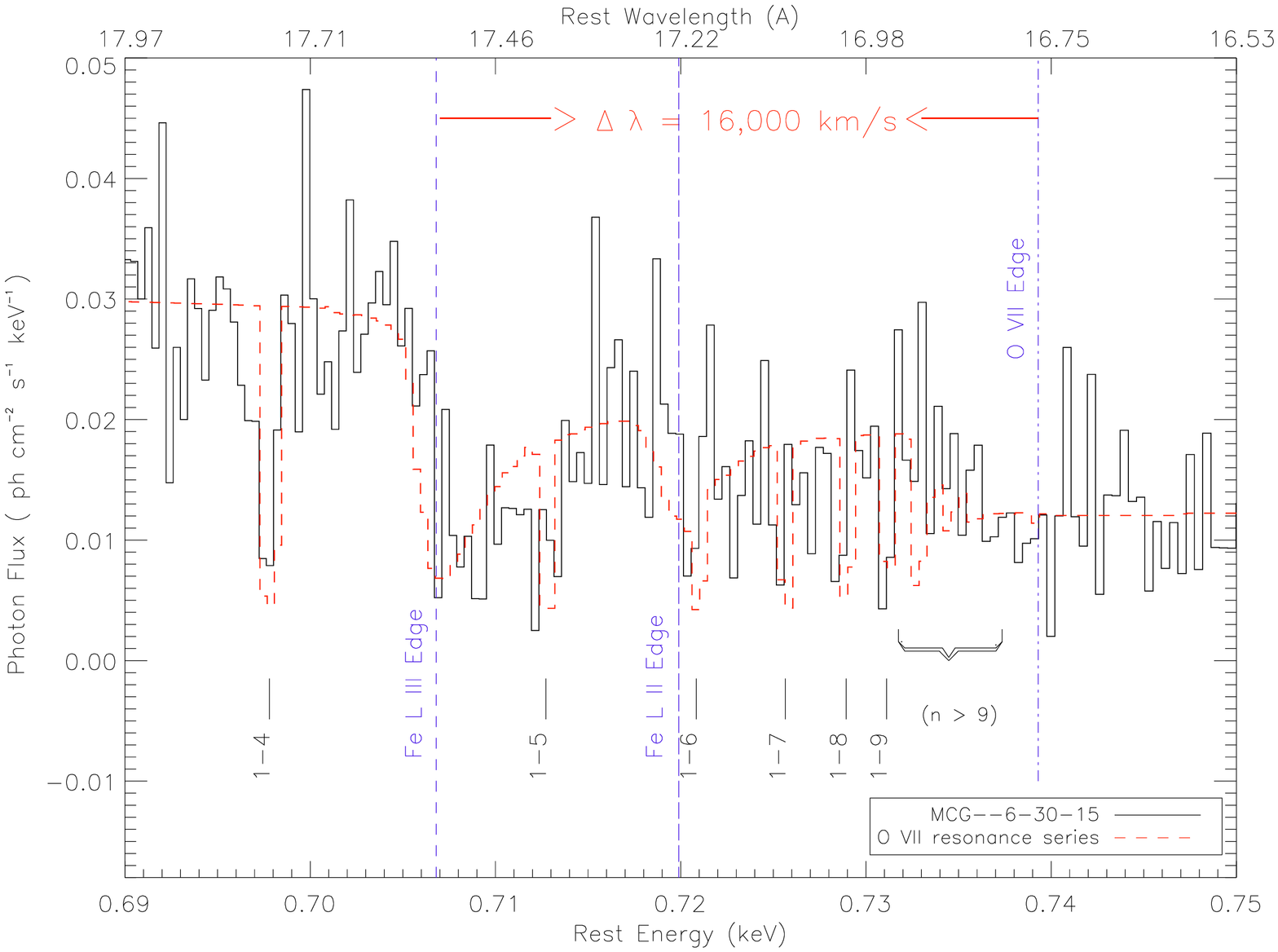,angle=90,width=3.5in,height=2.8in}
\vspace{-0.4in}
\caption{The region under contention.  The O~{\sc vii} edge at 
0.74~keV (16.75~\AA) appears to be redshifted by 16,000~km/s.
We argue that this effect is not due to broadened relativistic
emission lines.  Instead, this plot of the O~{\sc vii} resonance 
absorption series for $n > 3$ (red)
overplotted on MCG--6-30-15 (black) shows that the apparent
$16,000 \rm km \, s^{-1}$ shift of the O~{\sc vii} edge (from 0.74~keV
to 0.70~keV) is explained by the O~{\sc vii} resonance series carving
out this region, and a neutral Fe~L edge feature attributed to dust in the warm
absorber. The labels `1--$n$' correspond to the transitions (i.e.
$1{\rm s}^2-1{\rm s}n{\rm p}$) to ground state; e.g. 1--4 correspond to
$1{\rm s}^2-1{\rm s}4{\rm p}$, or He~$\gamma$. }\label{fig-mcg6}
\end{figure}

BR2001 concluded that their \xmm RGS observation of MCG$-$6-30-15 was
``physically and spectroscopically inconsistent'' with ``standard''
warm absorber models.  Our \chandra HETGS spectrum of this source shows 
that a dusty warm-absorber model is
not only adequate to describe the spectral 
features $\approxgt 0.48$~keV ($\approxlt 26$ \AA), the data {\it require} 
it.  We detail below the primary criticisms of the BR2001 model as 
presented in the Lee et al., (2001) paper.  We note that 
at the time of writing, the BR2001 model is the only model presented for the 
\xmm data of MCG--6-30-15 for which 
details are published or in preprint form. \\

\noindent
{\color[rgb]{1,0,0} (1) XMM :} {\it The thrust of the BR2001 argument is the
inconsistency between the apparent highly redshifted O~{\sc vii} 
(by $\sim 16,000\, km\,s^{-1}$) and O~{\sc viii}
edges and the absence of corresponding resonance absorption lines,
which would be expected for any but the most contrived kinematic
models. In order to reconcile this apparent inconsistency, they 
propose that the spectra of MCG--6-30-15 (and Mkn~766) be
explained by strong, highly  relativistically broadened Ly~$\alpha$
emission lines of  H-like O, N and C from the near vicinity of a Kerr
black hole on top of a much flatter continuum. } \\

\noindent {\color[rgb]{0,0,0}
{\color[rgb]{0,0,1} (1) Chandra : }Our \chandra data shows that 
the {\it apparent} highly redshifted O~{\sc vii} edge which prompted the BR2001
interpretation can be explained in the context of {\bf absorption from a warm absorber,
with dust}.  Fig.~\ref{fig-mcg6} shows that 
this region can be well explained by a neutral Fe~L edge with a significant 
(equivalent $N_{\rm H} \sim \rm 4 \times 10^{21} \, cm^{-2}$) column density 
in congruence with the overlapping 
O~{\sc vii} $1{\rm s}^2-1{\rm s}n{\rm p}$,  $n \ge 5$ series (i.e. He~$\delta$ to
the series limit) of  absorption lines
which carves away at the spectrum redward of the O~{\sc vii} edge (i.e. 
between $\sim$~0.70 and 0.74~keV).

The Fe~{\sc i}~L~{\sc iii} (and L~{\sc ii}) edge features are particularly significant.
The most plausible explanation for this feature is
absorption by dust that is embedded in the partially ionized material;
consequently the bulk of the gas is ionized.
Otherwise, for the column densities ($N_{\rm H} \sim \rm 4 \times 10^{21} \, cm^{-2}$)
implied by the drop across the 
$\sim 0.7$~keV Fe~L~{\sc iii} edge, the soft spectral features would be absorbed.
[We note that the Fe~L~{\sc iii} edge at
0.707~keV (17.5~\AA\,) and the associated Fe~L~{\sc ii} edge show 
structure similar to that seen seen in X-ray
binaries (e.g. X0614+011 and  Cyg~X-1, respectively \cite{paerels01,schulz01},
and measured recently in the laboratory \cite{kortright00})]. 
The presence of dust in the warm absorber of  MCG$-$6-30-15 
was already identified based on the optical reddening of 
E(B--V) = 0.61--1.09 \cite{csr97dust}, which implies
$N_H \approx  4-7 \times 10^{21} \rm \, cm^{-2}$.
Our independent
measurement based on the Fe~L absorption is consistent with this value.
Therefore, the presence of neutral
absorption is not only explained, it is required by the observed reddening.
We stress that because of the observed E(B--V)~$>$~3 reddening, dust is
a necessary component (not an option) to any model for MCG--6-30-15 
and similarly reddened AGN.
Certainly, this is not the first time that dust has been proposed 
as a component of the warm absorber in this and other similar AGN 
(e.g. \cite{brandt96,csr97dust}, and references therein).  
A neutral Fe:O ratio of 1:2  is 
supported by the {\it Chandra} data, and is consistent with dust composed of Fe oxides or 
fayalite ($\rm Fe_2SiO_4$), but see \S\ref{sec-toconsider}.}
\\

\noindent
{\color[rgb]{1,0,0} (2) XMM :} {\it The absorption lines detected by BR2001 are attributed to 
low ($N < 10^{17} \rm cm^{-2}$) column density
absorbers with velocity widths $\sim 2000 \rm \, km \,s^{-1}$ FWHM which 
they claim is insufficient to produce a detectable O~{\sc vii} edge. 
} \\

\noindent {\color[rgb]{0,0,0}
{\color[rgb]{0,0,1} (2) Chandra : }Indeed an O~{\sc vii} edge would not be detectable IF the O~{\sc vii} 
column densities are the low values quoted by BR2001.  However, Table~1 shows 
that we detect the O~{\sc vii} series from He~$\alpha$ ($n=1$)to He~$\eta$ ($n=9$).
The comparable values of the 
equivalent widths of these lines for oscillator strengths ($f_{ij}$) which 
differ by $\sim$~couple orders of magnitude (e.g. between the $n=1$ and $n=9$
O~{\sc vii} series) imply that they are on the flat part of the curve of growth 
(Table~1).  Accordingly, we performed a curve-of-growth analysis and estimate the 
minimum column density implied by the absorption lines to be 
$N_{O VII} \approxgt 7 \times
10^{17}  \, \rm cm^{-2}$ (for turbulent velocity width 
$b \sim 100 \rm \, km \, s^{-1}$) which requires an optical depth
$\tau_{O VII} > 0.2$  at the O~{\sc vii} edge.   This is consistent with the actual drop across 
the O~{\sc vii} edge at 0.74~keV (16.8~\AA\,) which gives $\tau_{O VII} \sim 0.6-0.8$, or $N_{\rm OVII}
\approx 2.5 \times 10^{18} \, \rm cm^{2}$.  These column densities are 
$\approx 1-2$~orders of magnitude greater than that quoted by BR2001.
These are not resolved in the \chandra HETGS.  We note
that the \chandra  resolution in this region is FWHM $\sim 0.023$~\AA \,
(compared to the \xmm resolution of $\sim 0.06$~\AA). 
A FWHM of 2000~km/s would correspond to a line width of  0.11\AA\, which we can
clearly rule out with the \chandra data.
 To summarize, the \chandra data 
show a lower limit of ($N_{\sc ovii} > 7 \times 10^{17} \rm cm^{-2}$) 
based on the detected O~{\sc vii} resonance lines 
(velocity widths $\approxlt 200 \rm \, km \,s^{-1}$ in contrast to the order of 
magnitude larger $\sim 2000 \rm \, km \,s^{-1}$
widths quoted by BR2001), and $N_{O VII} \sim 2.5 \times 10^{18} 
\rm cm^{-2}$ from the discontinuity at the $\sim 0.74$~keV O~{\sc vii} edge
(which is at zero velocity).   \\

\noindent
{\color[rgb]{1,0,0} (3) XMM :} {\it  The BR2001 model, as published, implies that 
the rollover in the continuum below $\sim
2$~keV is due to a $\Gamma \sim 2.0$ to $\Gamma \sim 1.3$ break in the power-law. 
We note, however, that as of this meeting our understanding is that this model 
is currently being revised.} \\

\noindent {\color[rgb]{0,0,0}
{\color[rgb]{0,0,1} (3) Chandra : }
Since we do not know the details of the revised model to explain the \xmm data,
we will address the published model as presented in the BR2001 paper. 
We note however that any model which allows for significant emission at the 
lower energies will require a suppression of the continuum.  
We believe the break implied in BR2001 is unphysical, especially
since such a flat continuum lying under the soft X-ray disk-line 
violates the Comptonization argument for the hard X-ray emission \cite{hm91}.
Furthermore, we have shown that the multi-component warm absorber 
introduces an accumulation of absorption lines and  edges (in particular 
various stages of Fe, Ne, Mg) which causes the spectrum to roll over 
below 2~keV (e.g. Fig.~\ref{fig-Gpl}).  An 
excess soft continuum (e.g. black-body or steep power-law) is required 
to compensate for this absorption at energies $<$~0.7~keV,
as seen in many other AGN.
If it were not for the absorption, a single power-law with Galactic
absorption could roughly account for the  flux below 0.7~keV and above
2~keV (Fig.~\ref{fig-Gpl}).   \\
}

\subsection{Additional points to consider} \label{sec-toconsider} 
\vspace{0.1in}
Of course, we cannot
rule out some small contribution from relativistically broadened
emission lines as well, and our data cannot address the shape of the
spectrum below 0.48~keV. However, since we 
see no evidence for such line emission and we can explain the great 
many features we do see
as coming from a dusty warm absorber, we conclude that 
the BR2001 interpretation of a spectrum 
dominated by relativistic line emission is unwarranted.
We note below some additional points to consider. \\

\noindent 
\noindent$\bullet$
A strong testament to the existence of the warm absorber is the myriad
of ionized species (e.g. Fig.~1a in Lee et al.).
There are also many absorption lines of higher ionization states of Fe up to Fe~{\sc
xxiii},  Mg, Si, S, Ar, and possibly Ca dispersed through the 0.9--5~keV
bandpass (Lee et al., in preparation).  \\

\noindent$\bullet$
Our \chandra spectra cannot be used to address the spectral features
below 0.48~keV ($> 26$~\AA).     Accordingly, we make
a rough estimate of the Fe:O ratio in dust -- while we believe it is
reasonable, it may not be entirely appropriate for modeling the low
energies (for e.g. the C absorption edge) to
which our instrument is not sensitive.   The exact shape of the soft 
excess will be important for properly modeling the composition of 
the dust in MCG--6-30-15, and needs further consideration.  
However, we stress that dust has to
be included in any model of MCG--6-30-15.  What remains is
to deduce from the data its properties, and composition. \\

\noindent$\bullet$
As stated in Lee et al., the detection of the O~{\sc vii}~(forbidden) emission
in the \chandra spectra would necessitate that re-emission
following resonance absorption can fill in some of the resonance lines,
thereby making the absorption appear weaker than it actually is.
It should be noted also that the amount of resonance re-emission in a 
clumpy warm absorber into our line of sight is entirely
dependent on the geometry.  This is well known from studies of stellar
coronae, including that of the Sun.  Accordingly, the scenario 
in which recombination cascades dominate over resonance re-emission 
will only apply for an isotropic absorber (uniformly filled sphere or slab)
which is clearly an unwarranted geometry as demonstrated by the 
calculation based on the O~{\sc vii}~(f) line in our paper. 
Photoexcitation was not addressed.

%do not change this

%do not change this
\small

\section*{ACKNOWLEDGEMENTS}
We thank Jeff Kortright, Anil Pradhan, Max Pettini, Eric Gullikson,
and many of our colleagues in the MIT HETG/CXC group, with
special thanks to Kathryn Flanagan.   We acknowledge the great efforts of the
many people who contributed to the \chandra program.  The work at MIT
was funded in part by contract SAO SV1-61010 and NASA contract
NAS8-39073.  ACF thanks the Royal Society for support.

\end{document}